\documentstyle[12pt,epsf]{article}

\newcommand{\k}{{\bf k}}
\newcommand{\prl}{ Phys. Rev. Lett.}
\newcommand{\prb}{ Phys. Rev. B }

\begin{document}

\title{Properties of the {$t$-$J$} Model in
the Dynamical Mean Field Theory:
One-Particle Properties in the Antiferromagnetic Phase}

\author{ Thomas Obermeier, Thomas Pruschke, Joachim Keller \\
{\it Institut f\"ur Theoretische Physik,
 Universit\"at Regensburg,
D-93040 Regensburg} \\
\\}

\date{November, 1995}

\maketitle
\begin{abstract}
We study the one-particle properties of the $t$-$J$ model within the framework
of Vollhardt's dynamical mean field theory. By introducing an $AB$ - sublattice
structure we explicitely allow for a broken symmetry for the spin degrees of
freedom and are thus able to calculate the one-particle spectral function in
the antiferromagnetic phase. We observe surprisingly rich structures in the
one-particle density of states for $T<T_N$ at finite doping up to 15\%.
These structures can be related to the well known results for one single hole
in the N\'eel background. We are thus able to establish
the relevance of this at a first sight academic limit to physical properties
of the $t$-$J$ model with a finite density of holes in the thermodynamical
limit.
\end{abstract}

\baselineskip 20pt
\pagebreak
\section{Introduction and Model}
 From the very beginning of the investigation on the properties of the
high-temperature superconductors \cite{bednorz} it was obvious that strong
local correlations together with long-ranged antiferromagnetic (AF) exchange
play an important role for the understanding of the anomalous
normal state properties of these materials \cite{anderson87}.
Several models of strongly correlated electrons
have been proposed since then to describe the
essential physics of the cuprates
\cite{dagotto}.
The most common ones are the one- and three-band Hubbard model \cite{HM,Emery}
and the $t$-$J$ model \cite{Zhang}.
The latter is the most basic one taking into account explicitely
both the strong local correlations and the magnetic interactions among
itinerant electrons. It can be viewed at as an
effective model describing essentially the
low energy physics of the more general
Hubbard models in the strong coupling limit
\cite{Chao77,Hirsch85,Zhang}.

Using standard notation, the Hamiltonian of the $t$-$J$ model reads
\begin{equation}\label{tJmodel}
H_{t-J} = -t \sum_{{\rm <ij>}\sigma} \tilde{c}^{\dagger}_{\rm i \sigma}
\tilde{c}_{\rm j \sigma} +
J \sum_{\rm <ij>}( \vec{S_{\rm{i}}} \cdot \vec{S_{\rm{j}}} - \frac{1}{4}
n_{\rm i} n_{\rm j}),
\end{equation}
where $\rm<ij>$ denotes the sum over nearest neighbours only
and $\tilde{c}^\dagger_{i\sigma}=(1-n_{i,-\sigma})c^\dagger_{i\sigma}$
creates an electron at site $i$ with spin $\sigma$ if and only if this site
is unoccupied.
Due to the constraint of no double occupancy imposed by the
projected single-particle operators the hopping term is already nontrivial
and contains strong correlation effects.
The second term describes the AF exchange coupling of spins at nearest
neighbour
sites, denoted by the operators $\vec{S_{\rm{i}}}$ and $\vec{S_{\rm{j}}}$
respectively.
The $n_{\rm i}n_{\rm j}$-term appears in the strong coupling expansion of the
Hubbard model. We will drop it because it does not directly affect the
magnetic properties we are
interested in.

There is a large amount of knowledge about this model for the very special
case of only one hole moving in an AF background
(see e.g.  \cite{Bri70,kane89,becker,igarashi,dagotto} and references
therein.)
This limiting case proved to be of special interest since it could be solved
exactly for $d=\infty$ \cite{Str92}.
The most important physical
aspect is that the moving hole feels a binding potential growing linearly with
the distance from its starting point due to the disturbance of the AF
background during its motion \cite{dagotto}. This linear potential leads to
a sequence of discrete poles as spectrum
for the one particle excitations with a distinct dependence
on $J/t$ \cite{kane89,Str92}. This rather simple picture
appears to remain valid even for finite-dimensional systems \cite{dagotto},
when transverse spin fluctuations
are important.
In order to compare with experiments it is equally important to study
the case of quasiparticle motion away from half filling at finite
temperatures. To our knowledge so far no results are available for
this case in the thermodynamic limit.

Here we want to study this situation in the framework of a
dynamical mean field theory (DMFT), which is known to become exact
in the limit of infinite spatial dimensions. This approach provides an
extension of the work of Strack and Vollhardt \cite{Str92}
for one hole to the
more general situation described above.
We find a multi-peak structure in the spectral functions that can be related
straight forwardly to the discrete spectrum of the one hole case.

The paper is organized as follows.
We give a short review of our method and notation in the next section.
In Sec.3 the possible evaluation of a magnetic
phase diagram is considered.
Sec.4 shows resulting spectral functions and
the first part of their interpretation. A brief review on the physics of one
hole in the fully polarized AF
state together with an interpretation of
the results from section 4 in connection with this picture
is then given in Sec.5, before the concluding remarks of Sec.6 summarize
the paper.

\section{Method}

The ideas of the DMFT for correlated electrons
on a lattice are based on the substantial simplifications found in the limit
of infinite dimensions \cite{Vol92}.
The most striking consequence of the $d$=$\infty$-limit concerns
the irreducible one-particle selfenergy $\Sigma_{\sigma}(\k,z)$,
which is due to the Coulomb interaction in Hubbard-like
models. This quantity becomes purely local, i.e.\ it looses all $\k$
dependence in the limit $d$=$\infty$ \cite{Met89,Mue89}.
Nevertheless it keeps a
highly nontrivial behaviour as a function of frequency.

In order to perform the limit $d\to\infty$ systematically
the hopping matrix element $t$ has to
be rescaled by
\begin{equation}
t = \frac{t^*}{2 \sqrt{d}}
\end{equation}
with a constant $t^*$
\cite{Met89}.
We will chose $t^* = 1$ as the unit of energy.
For a simple hypercubic lattice with nearest-neighbour
hopping only the bare density of states is then found to be a
Gaussian \cite{Mue89}:
\begin{equation}
\rho_0(\varepsilon) = \frac{1}{\sqrt{\pi}} e^{-\varepsilon^2}.
\end{equation}

Besides $t$, also the next neighbour exchange interaction
$J$ (which is proportional to $t^2/U$) has to be rescaled according to
\begin{equation}
J = \frac{J^*}{2d}\;\;.
\end{equation}

It is well known that this type of interaction becomes
trivial for $d\to\infty$ in the sense that the Hartree approximation
becomes exact \cite{itzykson,Mue89}.
In our case this means that the spin-interaction has to be
treated at the mean field level:
\begin{eqnarray}\label{MF}
J \sum_{\rm <ij>} \vec{S_{\rm{i}}} \cdot \vec{S_{\rm{j}}} &
	\longrightarrow &
 J \sum_{\rm <ij>} (  \vec{S_{\rm{i}}} \cdot <\vec{S_{\rm{j}}}>+
		   <\vec{S_{\rm{i}}}> \cdot \vec{S_{\rm{j}}}-
		   <\vec{S_{\rm{i}}}><\vec{S_{\rm{j}}}> ) \\
	 & \longrightarrow &
	2 J \sum_{\rm <i,j>} S_{\rm{i}}^z  <S_{\rm j}^z>,
\end{eqnarray}
for a polarization in $z$-direction.
Here $2<S_{\rm{j}}^z> = <n_{{\rm j}\uparrow} - n_{{\rm j}\downarrow}>$
\cite{comment1}.
Obviously $<S_{\rm{j}}^z>$ vanishes
in the paramagnetic phase and in this case the $t$-$J$ model for $d$=$\infty$
is equivalent to the $U$=$\infty$ Hubbard model.

Here we are interested mainly in the one-particle properties of
the $t$-$J$ model (\ref{tJmodel}) {\em in}\/ the antiferromagnetic state, i.e.\
when
$<S_{\rm{j}}^z>\ne0$. To allow for such a solution we must provide
a small symmetry-breaking field and in addition formulate all quantities
on an A-B lattice \cite{brandt}. Note that
the use of two sublattices does not affect the
locality of the self energies! Hence, the one-particle Green's function may
be represented as a matrix on the sublattices of the form
\begin{eqnarray}
G_{\sigma}(\k,z) =  \left( \begin{array}{cc}
	\zeta^{\rm A}_{\sigma}  & -\varepsilon_{\k} \\
	-\varepsilon_{\k} & \zeta^{\rm B}_{\sigma}
	\end{array}
	\right)^{-1} \;\;.
\end{eqnarray}
For simplicity we introduced the abbreviation $\zeta^{\rm A/B}_{\sigma}=
z+\mu - \Sigma^{\rm A/B}_{\sigma}(z) + h^{\rm A/B}_{\sigma}$
with the local selfenergies $\Sigma^{\rm A/B}_{\sigma}(z)$
and the staggered magnetic field
$h^{\rm A/B}_{\sigma} = \frac{1}{2}J^* \sigma
(<n_{\uparrow}^{\rm B/A}-
n_{\downarrow}^{\rm B/A}>)$, which results from the exchange interaction
(\ref{MF}) on a hypercubic lattice.

The local Green's function is obtained by summing over $\k$
with the result
\begin{equation}\label{Gloc}
 G^{\rm A/B}_{\sigma}(z) =
\int d\varepsilon \frac{\zeta^{\rm B/A}_{\sigma}}{\zeta^{\rm A}_{\sigma}
\zeta^{\rm B}_{\sigma} - \varepsilon^2} \;  \rho_0(\varepsilon) \;\;.
\end{equation}
Due to the additional symmetry \cite{brandt}
\begin{equation}
 G^{\rm A}_{\sigma}(z) = G^{\rm B}_{\overline{\sigma}}(z)
\end{equation}
it is sufficient to perform the calculations for the A-sublattice only and to
use the spin index for the bookkeeping.

The actual calculation is a straight forward extension of
the method used for the para\-mag\-netic phase \cite{Geo92,Pru..,Hue94}.
Due to the local nature of the selfenergies it is possible to
reduce the solution of the
lattice problem to that of a single site impurity problem
with local energy $(\varepsilon_{\sigma}-\mu)
=-(\mu + h_{\sigma})$
in contact with
an effective medium described by a Green's function
${\cal G}_{\sigma}(z)$. The latter contains all contributions
from the local correlations except for those at the impurity site, i.e.
\begin{equation}
{{\cal G}_{\sigma}}^{-1}(z) = {G_{\rm ii,\sigma}}^{-1}(z) +
\Sigma_{\sigma}(z).
\end{equation}
This equation relates the Green's function of the effective medium
to the local Green's function $G_{\rm ii,\sigma}(z)$ defined by (\ref{Gloc}).
The remaining atomic problem is solved by using the resolvent method
\cite{keiter70,bcw85} and the so-called Non Crossing Approximation
(NCA) \cite{bcw85,Pru..}, from which we obtain the
self energies of the effective impurity problem. Numerically
exact methods like e.g.\ Quantum Monte Carlo are not available for this
particular problem due to the fact that we actually have to solve an impurity
Anderson model with $U=\infty$ \cite{bulla}.
Thus, taking the impurity self energy as new
approximation for the local self energies of the
lattice problem (\ref{Gloc}), we are able to determine $\Sigma_{\sigma}(z)$
selfconsistently by iteration.
For details of this procedure see e.g. \cite{Pru..}.

\section{Phase diagram}
In the following we are interested in solutions in the AF state. Therefore
we choose slightly different starting conditions for up and down spins
in our iteration procedure and look for solutions with a finite
order parameter
$<S^z>$, which is non\-vanishing only in the AF phase.
There are of course other methods for obtaining the phase boundary,
such as calculating the free energies for the para- and AF-solutions or the
susceptibility of the $t$-$J$ model. A detailed discussion of the static
magnetic properties following from the susceptibility will be presented
in a forthcoming publication \cite{Pru-unpub}.

\begin{figure}[htb]
\begin{center}
\unitlength1cm
\begin{picture}(12,8)
\put(1,1){\epsfxsize=10cm\epsfbox{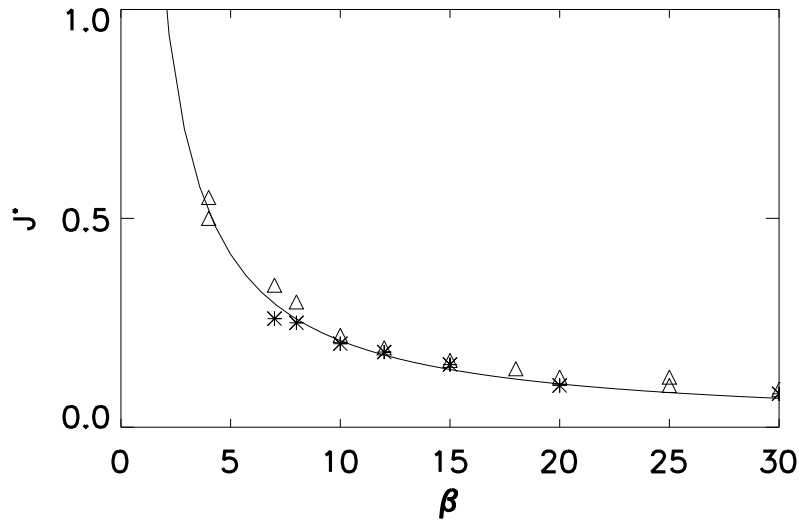}}
\end{picture}
\caption[]{Phase diagram $J_c(T,\delta)$ as function of $T$ for $n=0.98$.
The full line represents a fit to
the results from a calculation of the staggered
susceptibility \cite{Pru-unpub}.
These are compared with
our data points, which are marked by triangles if
$|n_{\uparrow}-n_{\downarrow}| > 0.0001$
and by stars otherwise.
}
\label{fig1}
\end{center}
\end{figure}

As it turns out, the task to calculate
the complete magnetic phase diagram of the $t$-$J$
model as function of temperature, doping, and $J/t$ using the method outlined
in the previous section cannot be performed for the following reasons.
First, the self-consistency cycle for the order parameter $<S^z>$
converges extremely slow close to the phase boundary.
Second, we could not set
a reasonable boundary for $<S^z>$ below
which one may safely assume that one is indeed in the paramagnetic phase
and not in the AF but with only a very small staggered magnetization. The
latter uncertainity for example leads to a comparatively large error in
the exact position of the phase boundary.
We thus rest content with comparing a few points of the phase diagram obtained
with the present calculations to the more accurate results obtained from the
staggered susceptibility \cite{Pru-unpub}.
This is done in Fig.~\ref{fig1}, where we show results for the critical $J$
as function of $T$ at fixed $n$. Evidently both methods lead within the error
bars to the same values of $J_c$, as was to be expected.

\section{Spectral functions}
\begin{figure}[htb]
\begin{center}
\unitlength1cm
\begin{picture}(12,8)
\put(1,1){\epsfxsize=10cm\epsfbox{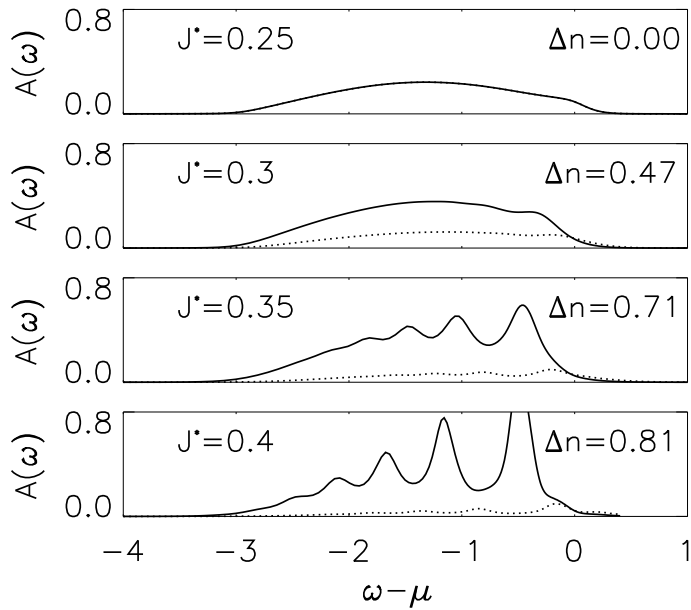}}
\end{picture}
\caption[]{Single-particle DOS for the $t$-$J$ model at $\beta=8/t^\ast$ and
$n=0.98$ for different values of $J^*$. The full curves represent the DOS for
the majority spin and the dashed curves the one for the minority spin on a
given site. Note that with increasing polarisation $\Delta n=
|n_{\uparrow}-n_{\downarrow}|$
the DOS develops pronounced structures below the Fermi energy.}
\label{fig2}
\end{center}
\end{figure}

We now concentrate on the behaviour of the spectral function
\begin{equation}
A_{\sigma}(\omega)=-\frac{1}{\pi}{\rm Im} G_{\sigma}(\omega + {\rm i}
	\delta)
\end{equation}
close to the transition from the paramagnetic to the AF state.
Fig.~\ref{fig2} shows the typical behaviour as the coupling constant $J^*$
is increased for fixed temperature $T=t^*/8$ and filling
$n=n_{\uparrow}+n_{\downarrow} = 0.98$.
For $J^*=0.25$ the system is in the paramagnetic state, in which case
the model is equivalent to a $U$=$\infty$ Hubbard model.
The structure near the chemical potential was earlier already related to some
kind of single-band Kondo effect \cite{Pru..,Pru_95}.
With increasing $J^*$ one eventually enters
into the AF regime, where the spectral
functions of up and down spins (full and dotted lines, respectively)
become inequivalent
and show a pronounced multi-peak structure. The latter will be discussed in
the next section.
A similar series of pictures would be obtained,
if the filling or the temperature were varied with the other parameters
fixed.

In order to show that the structures in both spin directions have the same
origin, we shift the energy scales about the staggered magnetic field
$h_{\sigma}$ that enters the local energy $\varepsilon_{\sigma}$.
Fig.3 shows that indeed the peak positions coincide.

Note that in the following figures we will use energy scales for the
different spectral functions, therefore,
that are shifted in the same way.
The spectral function of the minority spins (we
will refer to it as $A_{\downarrow}(\omega) $ in the following)
will sometimes
also be rescaled in height by a suitable factor in order
to be easier compared with  $A_{\uparrow}(\omega)$.
Note also, that the total spectral weight for the two  spin directions may be
different since we are looking only on the  lower  Hubbard band for
which no  sumrule applies.
\begin{figure}[htb]
\begin{center}
\unitlength1cm
\begin{picture}(12,9)
\put(1,1){\epsfxsize=10cm\epsfbox{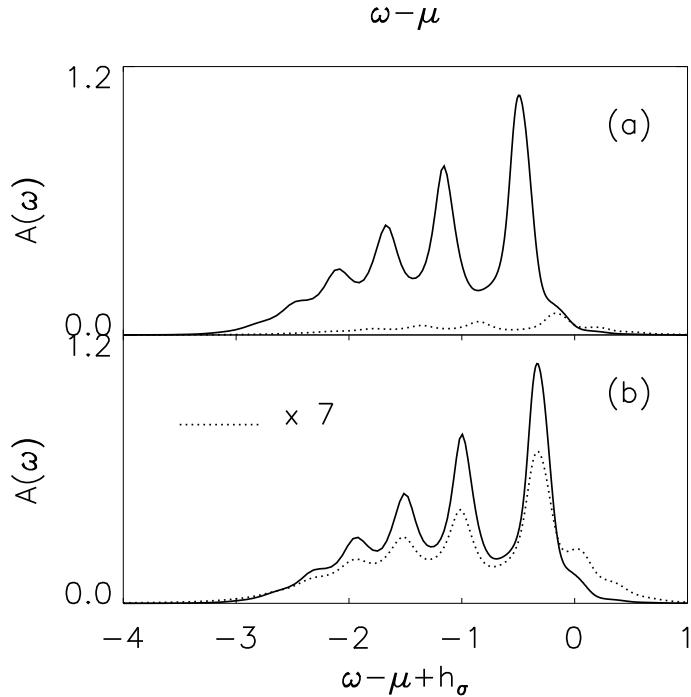}}
\end{picture}
\caption[]{(a) Single-particle DOS for $\beta=8/t^\ast$, $J^\ast=0.4t^\ast$ and
$<n>=0.98$ with $\Delta n=0.81$.
(b) DOS for the same parameters as in (a) but with the zero of energy
chosen according to the individual $\mu_\sigma$. Also, the minority DOS was
rescaled by a factor of $7$ to show the structures more clearly.}
\label{fig3}
\end{center}%
\end{figure}

An interesting problem concerns the question what happens to
the Kondo-like peak, observable in the paramagnetic state
(Fig.~\ref{fig2}) near the Fermi energy
\begin{figure}[htb]
\begin{center}
\unitlength1cm
\begin{picture}(12,8.5)
\put(1,1){\epsfxsize=10cm\epsfbox{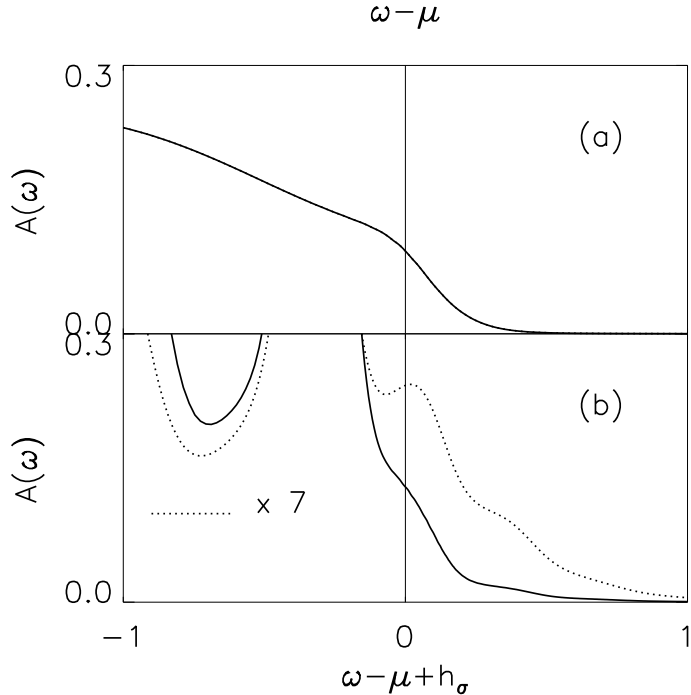}}
\end{picture}
\caption[]{(a) DOS for $J^\ast=0$, $\beta=8/t^\ast$ and $<n>=0.98$. Note the
structure at $\omega=\mu$ in $A(\omega)$. (b) DOS for $J^\ast=0.4t^\ast$,
other parameters
as in (a). The resonance at the Fermi level observed in (a) appears again at
the same position in both spectral functions. Since those were shifted by an
amount $h_{\uparrow}-h_{\downarrow}$ with respect to each other, this structure
is split in comparison to the case $J=0$.}
\label{fig4}
\end{center}%
\end{figure}
in the case of a finite sublattice magnetization.
In fact, for low temperatures and  small doping
the spectral function  shows an additional shoulder at the Fermi  energy if
the spectra are shifted by the molecular field  $h_\sigma$ (Fig. ~\ref{fig3}).
This can be seen clearly in Fig.~\ref{fig4}, where we  compare
the  spectrum in the antiferromagnetic state whith the corresponding
spectrum in the absence of the exchange interaction.
It is tempting to identify this feature with a Kondo resonance in the
antiferromagnetic state. However, due to the restricted parameter and
temperature range it is not possible to answer this question
definitely and to
exclude artefacts of the non-crossing approximation.

\section{Comparison to one hole in the N\'eel state}

We now want to discuss in detail the multi-peak structure in the
spectral function, which in our opinion can
be related to bound states of a hole in a
linear potential. This becomes clear if we compare our results with
the theory of Strack and Vollhardt \cite{Str92}. There it is shown
that the problem of one hole in the N\'eel background
can be solved exactly in the limit $d=\infty$.
In addition, a particularly comprehensive deduction of the one-particle
Green's function is given.

Before we start to present their results
we should mention, that the definition of the Green's
function and the choice of the energy scale in \cite{Str92} are slightly
different from ours: The zero of energy is the energy of the N\'eel state with
one hole $|{\rm i}N> = c_{\rm i}|N>$ and the hole's Green's function is
defined by
\begin{equation}\label{gf_hole_strack}
G^h_{\rm ii}(\omega) = <{\rm i}N| \frac{1}{\omega-\hat{\rm H}} |{\rm i}N>.
\end{equation}
 From this definition it can easily be seen that the sign of energy is
changed and the energy scale is shifted about the chemical potential
and the exchange field.
Note also that a different way of scaling for the hopping energy in
(\ref{tJmodel}) was used, namely
$t=\hat{t}/\sqrt{2d}$.
According to this we have to transform the results of \cite{Str92} in a
proper way before comparing them to our thermodynamic Green's functions.

Following the notation of \cite{Str92} the hole's Green's function reads
\begin{equation}\label{gf_hole_rpa}
  G^h_{\rm ii}(\omega) = \frac{1}{\displaystyle \omega
                -{\hat{t}}^2 \frac{1}{\displaystyle \omega
                - \frac{J^*}{2}-{\hat{t}}^2 \frac{1}{\displaystyle \omega
                - 2\frac{J^*}{2}-{\hat{t}}^2 \frac{1}{\displaystyle \omega
                - 3\frac{J^*}{2}-{\hat{t}}^2
			\frac{1}{\displaystyle \dots}}}}}\;\;.
\end{equation}
It has poles at frequencies
\begin{equation}\label{gf_hole_poles}
\omega_n = -2 \hat{t} - \frac{J^*}{2} - a_n t^* \left[ \frac{J^*}{2\hat{t}}
\right]^{\frac{2}{3}}\;\;,
\end{equation}
where $a_n$ are the zeros of the Airy function $Ai(4\hat{t}/J^*)$.
$G^h_{\rm ii}(\omega)$
 has been calculated using the retraceable path approximation of Brinkman
and Rice (BR) \cite{Bri70},
which becomes
exact in the limit of high dimensions \cite{Met92}.
The poles correspond to bound states of a particle in a linear or string
potential $\tilde{V}_m=m J^*/2$.
It is generated if the hole moves away from the
initial site by $m$ steps, destroying pairs of antiferromagnetic bonds
along its path, thus enhancing the energy
roughly by $m \times (2 E_{bond}) = m J^*/2$ (cf. Fig.5).
\begin{figure}[htb]
\begin{center}
\unitlength1cm
\begin{picture}(12,8)
\put(0,-1){\epsfxsize=14cm\epsfbox{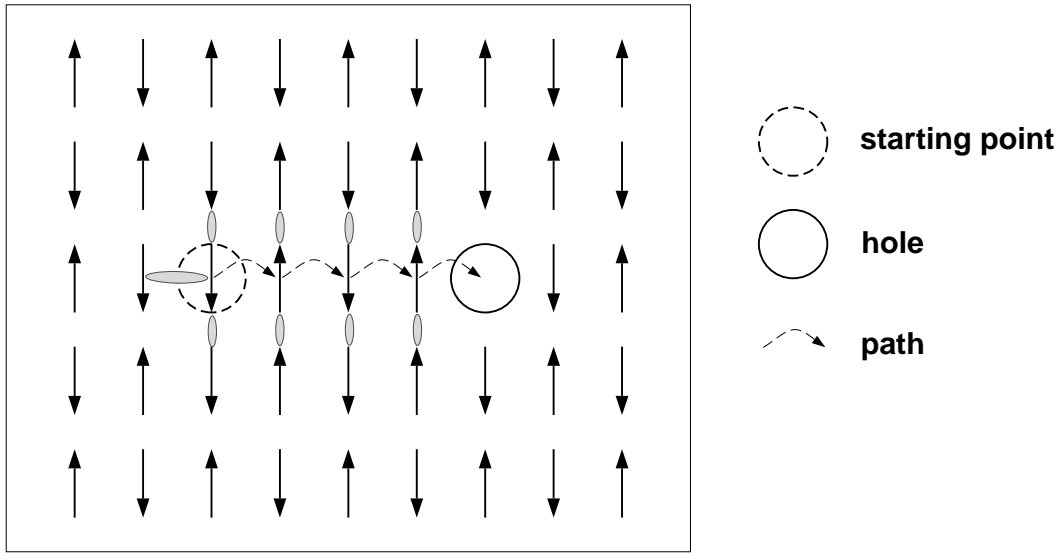}}
\end{picture}
\caption[]{Illustration of the origin of the linear potential for
a single hole moving in an antiferromagnetic background. The shaded
symbols mark destroyed bonds along the hole's path.}
\label{fig5}
\end{center}%
\end{figure}
Note, however, that even in $d$=$\infty$
the potential $ \tilde{V}_m$
is only exact and well defined for
the described situation of one hole moving in the AF background.
Therefore, the most important question that arises obviously is to
what extent these bound states will survive and how the spectrum will
be modified, as one turns
to low but finite doping and to finite temperatures.

Let us now return to our results for the one particle spectrum of the
$t$-$J$ model in the AF phase. In Fig.~\ref{fig3} a set of
distinct peaks can be seen,
which we believe are directly related to the bound
states of one hole. Of course, temperature and finite doping will lead to
the observed broadening. The more interesting question thus is to what
extent the positions of the peaks in Fig.~\ref{fig3} are related to
those in equation (\ref{gf_hole_poles}).

\begin{figure}[htb]
\begin{center}
\unitlength1cm
\begin{picture}(12,10)
\put(0,1){\epsfxsize=13cm\epsfbox{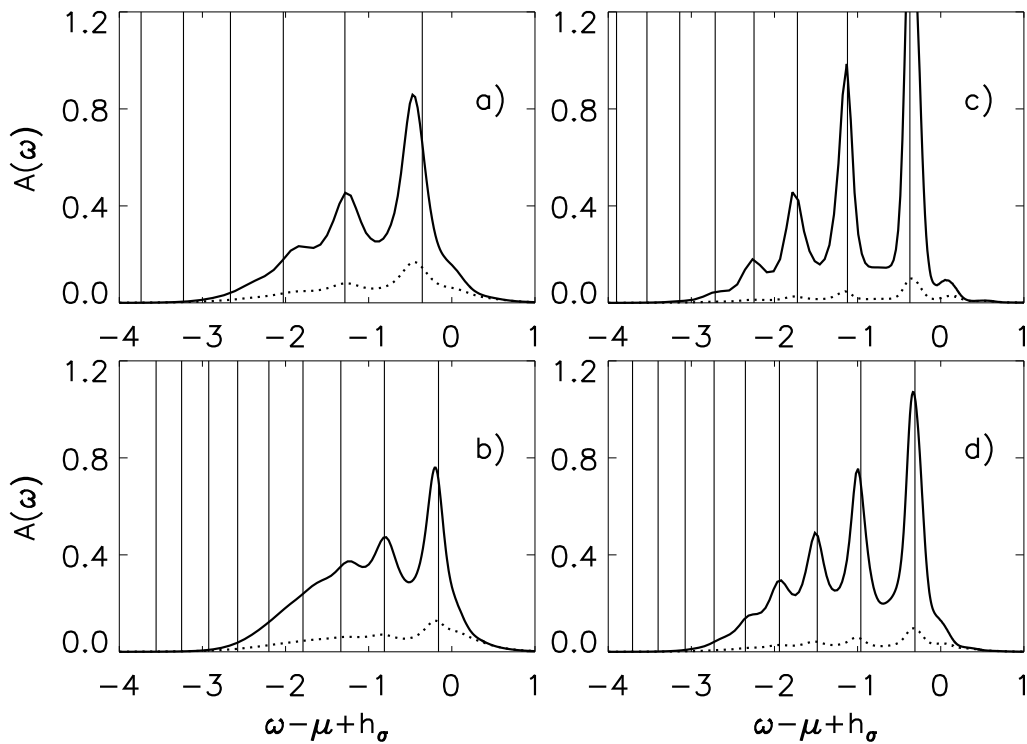}}
\end{picture}
\caption[]{Some examples for the fit of the peak positions by the modified
one-hole picture. Parameters are given in the text.
}

\label{fig6}
\end{center}%
\end{figure}

In Fig.~\ref{fig6} we present a direct comparison for the following
examples:
\newline
a) $J=0.7, n \approx 0.95, \Delta n \approx 0.64, \beta=4.0\;\;$
c) $J=0.5, n=0.98, \Delta n \approx 0.86, \beta = 7.0\;\;$
\newline
b) $J=0.4$, $n=0.95$, $\Delta n\approx 0.68$, $\beta = 8.0\;\;$
d) $J=0.4, n=0.98, \Delta n \approx 0.81, \beta=8.0\;\;$ .

The vertical lines represent the poles of the one-hole Green's
function (\ref{gf_hole_rpa}).
As mentioned before, the energy scale for the poles
has to be inverted and shifted to allow a direct comparison with
our data. As we cannot calculate accurately the chemical potential of the
N-particle N\'eel state, the shift is chosen to yield a reasonable
fit to the peak positions of $A_{\sigma}(\omega)$.
In the four examples we find quite good agreement
with the distance of the peak positions, particularly when
the system is very close to half filling like in the cases b) and c).
This also holds for other parameters we examined.

One can try to simulate the effect of finite doping in the theory
of Strack and Vollhardt by
assuming isolated holes, which move in a
partially polarized background.
The analytic form of the Green's function of this hole
remains the same as in (\ref{gf_hole_rpa}),
with the only difference that $J^*$ is replaced by an effective
exchange interaction
\begin{equation}
J^*_{eff} = |n_{\uparrow}-n_{\downarrow}|J^*
\end{equation}
which is reduced by means of the polarization of the actual system.
This procedure leads to a better correspondence for the cases a) and d)
but not for the other cases.

\section{Conclusion}
In this paper
we presented results for the dynamics of the $t-J$ model in the
antiferromagnetic state. These were obtained in the framework of the
dynamical mean field theory which becomes exact in the limit of infinite
dimensions (systems with many next neighbours).
The spectral function of single-particle excitations for
finite doping and finite temperature has features (a sequence of peaks)
which are similar to those found for
a single hole in an antiferromagnetic groundstate and which can be
understood by the motion of a hole in a string potential.
Our results show that this special case is relevant also for
finite hole concentrations and finite temperatures, at least for
$d\to\infty$. In this limit
we were able to neglect spin-fluctuations in the antiferromagnetic state,
which become more important in low dimensions.
Then also the concept of a string potential breaks down.
Due to this effect we expect a further broadening of the peak structure
in the spectral function.  Therefore it is not clear wether
these structures will also be present in
$d=2$ and $d=3$ for finite doping.
However, for one hole  exact diagonalisation studies \cite{eder}
for one hole in $d=2$
show structures which can be related to the $d$=$\infty$ spectrum.

\section*{Acknowledgements}
We wish to acknowledge helpfull discussions with
D.\ Vollhardt,
W.\ Metzner,
F.\ Gebhardt,
P.\ van Dongen,
G.\ Uhrig,
K.\ Becker,
N.\ Grewe,
F.\ Anders,
R.\ Eder,
and many others.
One of us (TP) would like to thank the Department of Physics at the
university of Cincinnati, where part of this work was done, for its
hospitality.
This work was supported by
the Deutsche Forschungsgemeinschaft grant number \mbox{Pr 298/3-1}.

\newpage



\begin{thebibliography}{99}
\bibitem{comment1} Note that on this level the
neglected last term of the Hamilonian (\ref{tJmodel}) reduces to
$-\frac{1}{4}n_{\rm i}\linebreak <n_{\rm j}>$, i.e.\ it simply leads to a shift
in the
chemical potential unless we allow for a charge density wave.
\end{thebibliography}
\end{document}